\documentclass{aa} 

\usepackage{tabularx}
\usepackage{graphicx}
\usepackage{txfonts}
\begin{document}

   \title{An idealized general circulation model for the atmospheric circulation on the ice giants}

   \author{I. Guendelman
          \inst{1}
          \and
          Y. Kaspi\inst{2}
          }

   \institute{Atmospheric and Oceanic Science Program, Princeton University,
              Princeton, NJ, USA\\
              \email{ig1245@princeton.edu}
         \and
             Department of Earth and Planetary Science, Weizmann Institute of Science, Rehovot, Israel\\
             \email{yohai.kaspi@weizmann.ac.il}
}
   \date{}

  \abstract{ 
Uranus and Neptune are the least explored planets in the Solar System.  A key question regarding the two planets is the similarity of their observed flows despite the great differences in their obliquity and internal heating. To answer this fundamental question and understand the ice giants atmospheric circulation, we developed a new general circulation model (GCM). This tool will also be key to facilitating the success of future missions to the ice giants, for which atmospheric flows will be a measurable quantity. Past GCMs for the ice giants have struggled to reproduce the observed winds on Uranus and Neptune. Using our idealized GCM, we systematically explored how the zonal wind and meridional circulation respond to different model and physical parameters;  our main focus was on the depth of the domain. We show that in cases where the bottom layer of the model is deep enough, the simulated flow is independent of the meridional structure of the forcing temperature, indicating that dynamical processes, and not the imposed thermal forcing, are the dominant drivers of the circulation and the thermal structure. A momentum balance analysis further shows that meridional and vertical eddy momentum flux convergence are both central to maintaining the circulation. These results provide a physical explanation for the similarity of the flow on Uranus and Neptune although their solar and internal forcing are significantly different. The modeling framework developed in this study can serve as a foundation for the development of more comprehensive GCMs of the ice giants and help guide the interpretation of future mission data.}

   \maketitle
%

\section{Introduction}

\begin{figure}[htb!]
    \begin{center}
        \includegraphics[width=0.95\linewidth]{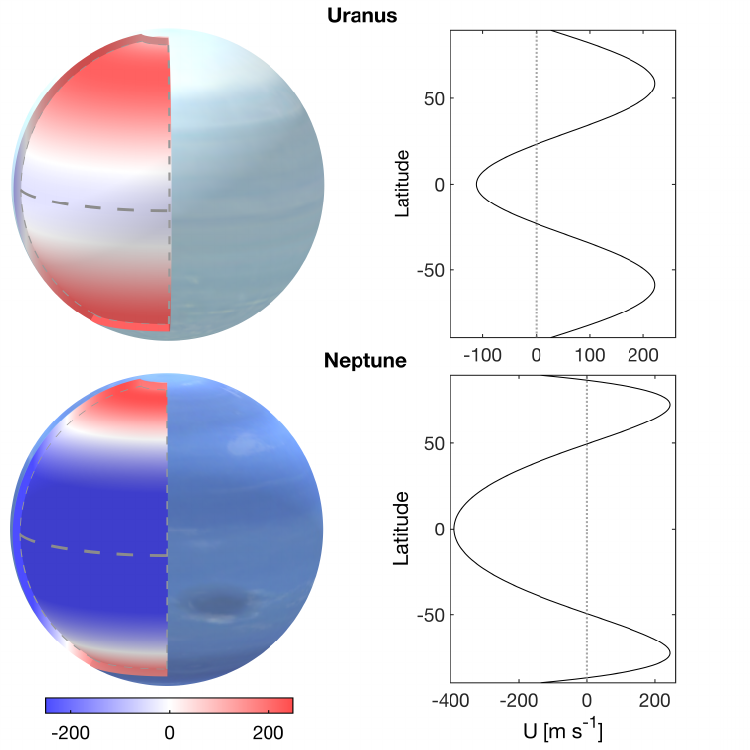}
    \end{center}
    \caption{Left: Cylindrical projection of the fit to the observed winds down to 0.95 of the planet radii for Uranus (top) and Neptune (bottom). Right: Analytical fit to the observed winds on Uranus \citep[top,][]{sromovsky_high_2015} and Neptune \citep[bottom,][]{french_neptunes_1998}.}
    \label{fig:fig_obs}
\end{figure}

The ice giants, Uranus and Neptune, are the least explored planets in our Solar System; one flyby of these planets was performed by Voyager 2 in 1986 and 1989, respectively \citep{flasar_voyager_1987,pearl_albedo_1990,pearl_albedo_1991}. Almost 40 years later, following successful missions exploring Saturn and Jupiter, there is increasing attention toward an orbiter mission to these planets \citep[e.g.,][]{ferri_atmospheric_2020,fletcher_ice_2020,dahl_atmospheric_2024}. Having scarce observations of these planets makes them difficult to study. A main example of the challenges of studying these planets is interpreting the atmospheric flow on these planets. Although both planets have very different internal and solar forcing \citep[e.g.,][]{helled_uranus_2020}, observations suggest that the cloud-level zonal wind structure on these planets is qualitatively similar, with a retrograde flow around the equator and a prograde flow in each hemisphere centered in the midlatitudes \citep[Figure~\ref{fig:fig_obs}, e.g.,][]{hueso_atmospheric_2019}. The similarity in the observed winds between Uranus and Neptune is puzzling given the different solar forcing (due to their different obliquity) and different internal heating. This observed flow is fundamentally different from the flow observed on the gas giants where there is a broad prograde equatorial jet and multiple jets poleward of it. An additional difference from the gas giants is that in the gas giants the observed banding in the atmosphere is correlated with the different jets \citep{duer_evidence_2021}, while this is not the case in the ice giants \citep{de_pater_neptunes_2014,karkoschka_uranus_2015,fletcher_ice_2020}. This difference might suggest that the ice giants observations are biased or that the relationship between the overturning circulation that is assumed to control the banding and the jets on the ice giants is qualitatively different from the gas giants. 

\cite{fletcher_ice_2020} suggested a qualitative description of the meridional circulation on Uranus and Neptune based on existing observations. However, there is a lack of both observations and modeling work on the ice giants. Only a few efforts have been made to simulate the general circulation of the atmosphere of the ice giants in a global circulation model. Most of these past efforts were part of efforts to simulate the flow on Jupiter and Saturn, and not to directly simulate the circulation of the ice giants. These efforts are generally divided into the use of deep or shallow circulation models, the difference between these efforts being the assumed depth of the atmosphere. In both approaches, the ability to produce a flow that resembles the observed flow is limited. \cite{lian_generation_2010} and \cite{liu_mechanisms_2010} showed in an idealized shallow model a transition from prograde to retrograde equatorial flow using different physical mechanisms; however, the simulated winds there were significantly weaker compared to the observed winds. \cite{aurnou_effects_2007} showed a transition in a deep model from a flow that resembles the one observed on the gas giants to a flow that resembles that observed on the ice giants; however, such models are driven with unrealistic internal heat fluxes, and thus cannot be compared quantitatively to the observed circulation.

The choice of the circulation model depends on our understanding of how deep the flow is in these planets. \cite{kaspi_atmospheric_2013} analyzed Voyager 2 gravity measurements and found an upper limit of $\sim1000$ km for the depth of the winds. This means that the circulation is limited to the upper $4\%$ or less of the planet, similarly to the case of Jupiter \citep{kaspi_jupiters_2018}. \cite{soyuer_zonal_2023} revisited this analysis and found a similar estimate for the depth. Similar estimations for the penetration depth of the winds are supported by magnetic considerations \citep{soyuer_constraining_2020}. These results suggest that a relatively shallow atmosphere model should be sufficient to reproduce the flow on these planets. However, studies using shallow models have not been able to do so. \cite{liu_mechanisms_2010} used an idealized model with idealized radiation, dry convection, and forced by annual mean insolation and idealized internal heating. Although they were able to generate a retrograde equatorial jet and approximately one broad jet in the midlatitudes, the winds produced in this model were significantly weaker compared to the observed ones. More recently \cite{milcareck_zonal_2024} have used a more complex model that incorporates different parameterizations of physical processes and a realistic radiation scheme with a seasonal cycle. Their results also suffer from weak winds, in addition to having a very complex zonal wind vertical structure that does not fit the observed winds.

Having a model that can simulate the flow on Uranus and Neptune is important for the ability to plan future missions and to interpret current and future observations. The challenges in doing so led to this work in which we employ an idealized general circulation model to simulate the atmospheric circulation on the ice giants. In Section~\ref{sec:model} we present and motivate the use of the model. In Section~\ref{sec:res} we present our results. We discuss their meaning for future simulation efforts in Section~\ref{sec:diss}.

\section{Model}\label{sec:model}

We used an idealized GCM that is forced by Newtonian relaxation to an equilibrium temperature \citep[e.g.,][]{held_proposal_1994},
\begin{equation}
    Q=-\frac{T-T_{\rm{eq}}(\phi,\sigma)}{\tau_r},
\end{equation}
where Q is the thermal forcing; $T$ is the temperature; $T_{\rm{eq}}$ is the radiative equilibrium temperature that the model is nudged toward, and that depends on latitude ($\phi$) and the vertical coordinate $\sigma=p/p_s$, where $p$ is pressure and $p_s$ is the pressure of the bottom layer of the model; and $\tau_r$ is the thermal relaxation timescale. Traditionally, $\tau_r$ is also dependent on $\phi$ and $\sigma$; for simplicity, we consider it to be constant both horizontally and vertically. We used the physical parameters of Uranus, for example rotation rate, radius, and gas constant, in the model simulations (Table~\ref{tab:parameters}). This model uses hyper-diffusion for numerical stability \citep{held_proposal_1994}.

The boundary layer friction in this model is a simple linear friction in the form of
\begin{equation}
    F=-K_u(\sigma)\textbf{u},
\end{equation}
where $\textbf{u}$ are the horizontal winds and 
\begin{equation}
    K_u = \tau_f\max\left(0,\frac{\sigma-\sigma_b}{1-\sigma_b}\right),
\end{equation}
where $K_u$ is the frictional timescale, $\tau_f$ is a constant, and $\sigma_b$ is the depth of the boundary layer. We used different values of $\sigma_b$ for different depths to keep the boundary layer relatively shallow in terms of atmospheric mass, and we tested the sensitivity to this parameter.

\begin{table}[]
    \caption{Physical and model parameters}
    \begin{center}
    \begin{tabularx}{\linewidth}{lr}
    \underline{\textbf{Physical parameters}}\\
        Planetary Radius $a$ ($10^6$ m) & 25.27 (24.55) \\
        Rotation rate $\Omega$ ($10^{-4}$ s) & 1.0124 (1.0834) \\
        Specific gas constant $R$ (J kg$^{-1}$ K$^{-1}$) & 3149.2 (3197.7)  \\
        Specific heat capacity $C_p$ ($10^4$ J kg$^{-1}$ K$^{-1}$) & 1.10 (1.12)  
    \\
    \underline{\textbf{Model parameters}}\\
        Thermal relaxation time scale $\tau_r$ (day) & 100 (1000) \\
        Frictional time scale $\tau_f$ (day$^{-1}$) & $100^{-1}$ ($1000^{-1}$) \\
        depth of the boundary layer $\sigma_b$ & $0.97$ ($0.85$) \\
    \end{tabularx}
    \\
    \end{center}
    {\small{
    \textbf{Notes.} The physical parameters are for Uranus and the values in parentheses are for Neptune, given to justify the use of the Uranus parameters in this study. For the model parameters the values in parentheses represent values for the sensitivity testing presented in Figure~\ref{fig:figusims}. The value given for $\sigma_b$ is for the 30 and 10 bar simulations, for the 1 and 3 simulation we used 0.8 and 0.87, respectively to keep a shallow boundary layer. The values for the physical parameters are from \cite{ladders_planetary_1998}.}} 
    \label{tab:parameters}
\end{table}

\begin{figure}[htb!]
    \begin{center}
        \includegraphics[width=0.95\linewidth]{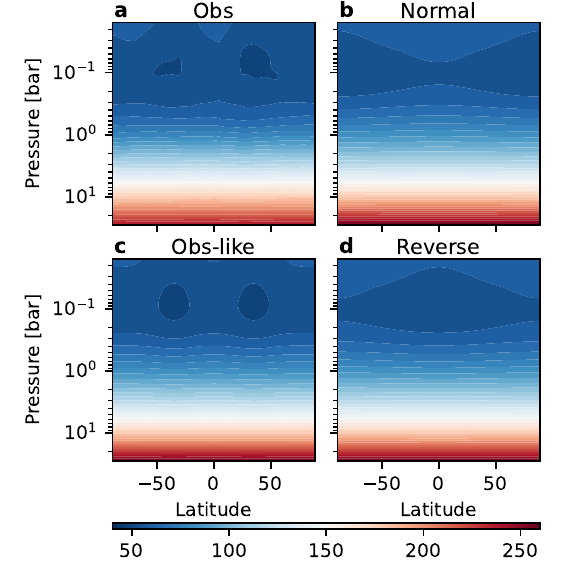}
    \end{center}
    \caption{Different equilibrium temperatures used to force the GCM. $T_{\rm{Obs}}$ is adopted from the Voyager 2 observations \citep{orton_thermal_2015}, $T_{\rm{Obs-like}}$ is a simplified fit to the observations, and $T_{\rm{Normal}}$ and $T_{\rm{Reverse}}$ are manipulations of the fit ($T_{\rm{Obs-like}}$) to obtain the forcing temperature maximum at the equator and poles, respectively.}
    \label{fig:Teq}
\end{figure}

We used different $T_{\rm{eq}}$ to force the model in different ways. We began by using the observed temperature for Uranus from the Voyager mission \citep[Figure~\ref{fig:Teq},][]{fletcher_neptune_2014,orton_thermal_2015}. We chose to use the temperature field for Uranus because the data are more spatially complete, whereas for Neptune there is a lack of observations poleward of $50^{\circ}$N. That said, if we assume that Neptune is close to being hemispherically symmetric, which the observations point to, the qualitative structures of the two planets are similar. The Voyager temperature profile, $T_{\rm{obs}}$, goes down to $\sim10$ bar; we extended it adiabatically down to $\sim30$ bar (shown in the bottom right panel in Figure~\ref{fig:Teq}a). We note that the confidence in the observed temperature profile is highest for the top of the atmosphere, i.e., for pressures less than $\sim 1$ bar and for deeper pressures the confidence is low, and at deeper levels the vertical temperature profile is close to the dry adiabat. We forced the model by using different depths of $T_{\rm{Obs}}$; more specifically, we chose 1, 3, 10, and 30 bar. Given the similarity in the temperature profiles, physical parameters (Table~\ref{tab:parameters}), and the use of Newtonian relaxation, we conducted our simulations using only the Uranus parameters, but the results are relevant for both planets.

Previous studies have used Newtonian relaxation combined with iterative techniques that are constructed with the goal of the simulated temperature to be similar to the forced temperature for both Earth and Mars \citep[e.g.,][]{chang_impact_2005,zurita-gotor_generalized_2006,zurita-gotor_relation_2007,yuval_effect_2017,mooring_effects_2019,yuval_eddy_2020}. While at face value a similar approach might be relevant here as well, we did not apply it in this study. This is for two main reasons. The first is that this study is more exploratory in nature, with the aim of identifying leading-order parameters and characteristics that influence the simulated atmospheric circulation; thus, our focus is less on the ability of the model to quantitatively replicate the observed flow, and more on simulating a qualitatively similar flow. The second reason for not using these techniques is that unlike Earth and Mars, where observations are abundant, for the ice giants we rely only on one observed profile and do not have other observations, which in turn limits our ability to confirm our simulated circulation. As already discussed, the confidence in the Voyager temperature profile is limited to the upper atmosphere, making the rationale for using these iterative methods obsolete. Additionally, relying on one observation of the temperature field increases the chance of it being biased, so the benefit of simulating the exact observed temperature profile is questionable. 

To test the sensitivity of our results to the detailed structure of $T_{\rm{eq}}$ we fit a simplified function to $T_{\rm{obs}}$ of the form
\begin{equation}\label{eq:fit}
    T_{\rm{fit}}=T_v(\sigma)+\Delta T(\phi,\sigma),
\end{equation}
where $T_v$ is representative of the mean vertical structure of the temperature, and $\Delta T$ represents the meridional structure of the temperature. The functional forms are detailed in Appendix~\ref{app:a}. In this case, the vertical structure of the meridional temperature gradient is simplified, which allows us to test the sensitivity of the resulting circulation to the detailed structure of the forcing temperature.

\begin{table}[htb!]
    \caption{Resolution details for each depth.}
    \begin{center}
    \begin{tabularx}{\linewidth}{llll}
        $p_s$ & Horizontal Resolution & Vertical levels & Timestep [s] \\ 
        1 bar & T170 ($\sim 0.7^{\circ}\times0.7^{\circ}$) & 30 & 1200 \\
        3 bar & T127 ($\sim 1^{\circ}\times1^{\circ}$) & 35 & 300 \\
        10 bar & T127 ($\sim 1^{\circ}\times1^{\circ}$) & 45 & 200 \\
        30 bar & T85 ($\sim 1.4^{\circ}\times1.4^{\circ}$) & 55 & 200 \\
    \end{tabularx}
    \end{center}
    
    \label{tab:res}
\end{table}

Following that, we conducted an additional sensitivity test in which we quantitatively changed the meridional structure of $T_{\rm{fit}}$, where in one case the temperature maximizes at the equator and in the second it maximizes at the poles. This test was motivated by the difference in obliquity between Uranus and Neptune. If we assume that the annual mean insolation is the leading-order forcing in both planets, and the seasonal cycle can be neglected, the difference in obliquity will result in the incoming solar radiation having a different latitudinal structure \citep[e.g.,][]{guendelman_key_2022}. For Neptune, the annual mean incoming solar radiation will maximize at the equator, while for Uranus, it will maximize at the poles. Another motivation for this sensitivity test comes from a recent study that used a radiative convective equilibrium model for the ice giants, where some of their resulting temperature profiles seem to be different from that observed by Voyager \citep{milcareck_radiative-convective_2024}. The results of this study might point to the role of dynamics in shaping the meridional temperature structure and motivate the examination of the circulation dependence on the meridional structure of the forcing temperature. 

The resolutions used in the different simulations are detailed in Table~\ref{tab:res}. All the simulations run to a point where the winds are in a quasi-steady state (see some examples in Figure~\ref{fig:conv} in Appendix~\ref{app:b}), and the results shown are a time mean over the last 2000 simulated days. We discuss the resolution dependence for different depths later in the manuscript; the different dynamical timesteps were chosen for the simulation stability.

\section{Results}\label{sec:res}
\subsection{Circulation dependence on depth}

\begin{figure*}[htb!]
\sidecaption
        \includegraphics[width=12cm]{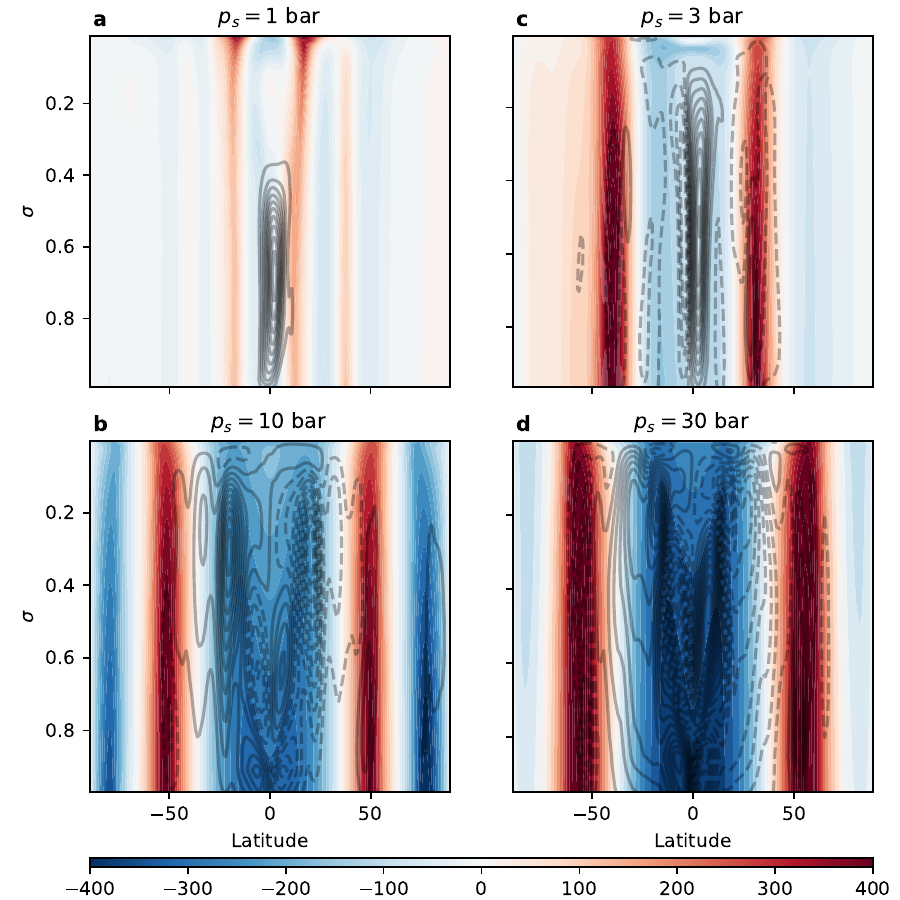}
    \caption{Zonal mean zonal winds  (m s$^{-1}$, shading), and the zonal mean streamfunction (contours; solid are for clockwise circulation, kg s$^{-1}$). The contour spacing for the streamfunction is $\max(\psi)/10$, for simulations forced by $T_{\rm{Obs}}$ with depths of 1, 3, 10, and 30 bar (panels a-d, respectively). For the 1 and 3 bar simulations (panels a and b) the winds are multiplied by 2 and 4, respectively, so that all panels share a common colorbar.}
    \label{fig:sim}
\end{figure*}

Figure \ref{fig:sim} shows the simulated circulation forced by $T_{\rm{Obs}}$ with different bottom-layer pressures taken from the extended observed temperature (Figure~\ref{fig:Teq}a). That is, $T_{\rm{eq}}=T_{\rm{Obs}}$ up to 1, 3, 10, and 30 bar (Figure~\ref{fig:sim}). The simulations that produced the zonal flow closest to the observed flow are the ones extended down to 10 and 30 bar (Figures~\ref{fig:sim} and \ref{fig:figusims});  the main difference between the two is a stronger polar retrograde jet in the 10 bar simulation. Generally, using a deeper bottom pressure, i.e., extending $T_{\rm{eq}}$ deeper, the zonal mean zonal winds strengthen (shading in Figure~\ref{fig:sim} and \ref{fig:figusims}a). In addition to the intensification of the zonal mean zonal winds, the jets shift poleward, the equatorial retrograde jet becomes wider, and the number of jets reduces with a deeper atmosphere. An additional difference between the simulations occurs when we consider the zonal mean meridional streamfunction, $\psi=2\pi a\int \overline{v}\cos\phi dp/g$, where $a$ is the planetary radius, $\overline{v}$ is the zonal mean meridional wind, and $g$ is the surface gravity (contours in Figure~\ref{fig:sim}; the solid contours represent clockwise circulation). For the 1 and 3 bar simulations, the circulation in the tropical latitudes (around the equator) consists of a narrow Hadley cell-like circulation, i.e., the air in the deeper levels near the equator rises and flows poleward to higher latitudes (Figure~\ref{fig:sim}). The width and symmetry of the circulation depends on the depth of the atmosphere, where the width is restricted to around $\sim 15$ degrees latitude from each side of the equator. Outside of this tropical circulation, the streamfunction is weak (represented by the lack of contours outside the tropical region in Figures~\ref{fig:sim}). This circulation pattern changes when the bottom-layer pressure is extended to 10 and 30 bar. In these cases, the circulation is more complex, consisting of stacked cells near the equator, where in the upper atmosphere (above $\sigma\sim0.2\ (0.4)$, i.e., $\sim6\ (4)$ bar in the 30 (10) bar simulation) the air descends at the equator and ascends in the midlatitudes, having some similarity to the circulation proposed by \cite{fletcher_ice_2020}. In addition, this complex circulation can align with the observed banding not being correlated with the observed winds \citep{fletcher_neptune_2014,orton_thermal_2015}. 

\begin{figure*}[htb!]
    \begin{center}
        \includegraphics[width=0.95\linewidth]{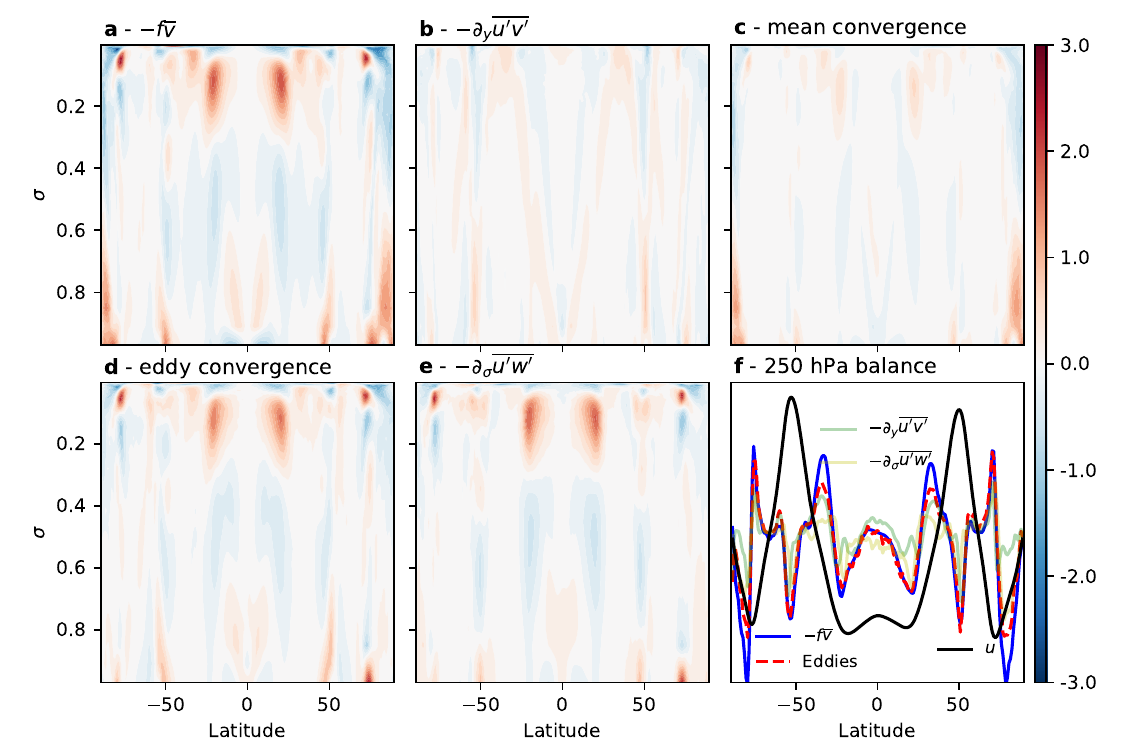}
    \end{center}
    \caption{Zonal momentum balance components for the 10 bar simulation (colorbar is in units of $10^{-4}$ m s$^{-2}$). (a) Coriolis acceleration $-f\overline{v}$, where $f$ is the Coriolis parameter. (b) Eddy meridional momentum flux convergence $-\partial_y\overline{u'v'}$. (c) Mean momentum flux convergence,  including the mean meridional flux of relative vorticity  $\overline{\zeta}\overline{v}$, where $\zeta$ is the relative vorticity, and the mean vertical momentum flux convergence $\overline{w}\partial_{\sigma}\overline{u}$. (d) Sum of eddy flux convergence, i.e., $-\partial_y\overline{u'v'}-\partial_{\sigma}\overline{u'w'}$. (e) Eddy vertical momentum flux convergence $-\partial_{\sigma}\overline{u'w'}$. (f) Leading order balance at 250 hPa. The vertical axis for the momentum budget terms is $\pm2\times10^{-4}$ m s$^{-2}$ and for the zonal mean (black curve) $\pm320$ m s$^{-1}$.}
    \label{fig:figmom}
\end{figure*}

When comparing the  simulated zonal mean zonal winds with the observations, the most comparable one is the 30 bar simulation;  the 10 bar simulation is a close second (Figure~\ref{fig:sim}c-d and \ref{fig:figusims}a). However, these simulations have significant differences from the fit to the observed winds. First, the winds are too strong  in the equatorial and in the midlatitude regions. Second, the width of the midlatitude prograde flow is too narrow compared to that suggested by the observations. Finally, in our simulations it seems that there is a subrotating jet in the polar latitudes, which does not appear in the observations. Although the fit of some of these flow properties can potentially be improved by tuning different model parameters, for example changing the boundary layer friction parameters (green curve in Figure~\ref{fig:figusims}b), given the limited observations, we find that there is no additional benefit to tuning the model parameters to get a better fit. Additionally, it is important to note two points when comparing simulated winds to observed ones. First, for this study we used a very idealized setting with the goal of simulating a circulation that is qualitatively similar to the observed one. Second, the wind observations are limited, and the possibility exists that future observations will show a more detailed structure of the winds, and that some new observations might show some contrasting results in the polar latitude of Uranus \citep{sromovsky_puzzling_2024}. Both of these reasons reduce the motivation to tune the model to best fit the observations.

The momentum balance can indicate the processes that are important for the maintenance of the meridional circulation and the winds. When considering the zonal momentum balance, it is customary to decompose the momentum fluxes into the zonal mean and eddy momentum fluxes; the zonal mean is denoted by a bar and the eddies are denoted by a prime. On Earth, the midlatitude circulation, i.e., the Ferrel cell, is eddy-driven, which means that  the Coriolis acceleration $-f\overline{v}$ at the top of the cell, where $f$ is the Coriolis parameter, is balanced by the meridional momentum flux convergence $-\partial_y\overline{u'v'}$ \citep[e.g.,][]{vallis_atmospheric_2017}. A similar balance was recently observed for Jupiter from the analysis of the microwave radiometer measurements by the Juno spacecraft \citep{duer_evidence_2021}. However, the leading order balance can vary depending on the planetary parameters \citep{guendelman_emergence_2021}. Figure~\ref{fig:figmom} shows the different contributions of the different components of the momentum fluxes in terms of the momentum convergence for the 10 bar simulation (see figure~
\ref{fig:app_30} for the 30 bar simulation). The main balance here is between the eddy convergence and the Coriolis acceleration (Figure~\ref{fig:figmom}a, d, and f); the vertical and meridional eddy momentum fluxes are both important, similarly to what was found in simulations of the gas giants \citep{duer_gas_2023}. A similar leading order balance is observed in the simulations with a more idealized $T_{\rm{eq}}$.

\subsection{Dependence on the meridional structure of $T_{\rm{eq}}$}

\begin{figure*}[htb!]
\sidecaption
    \includegraphics[width=12cm]{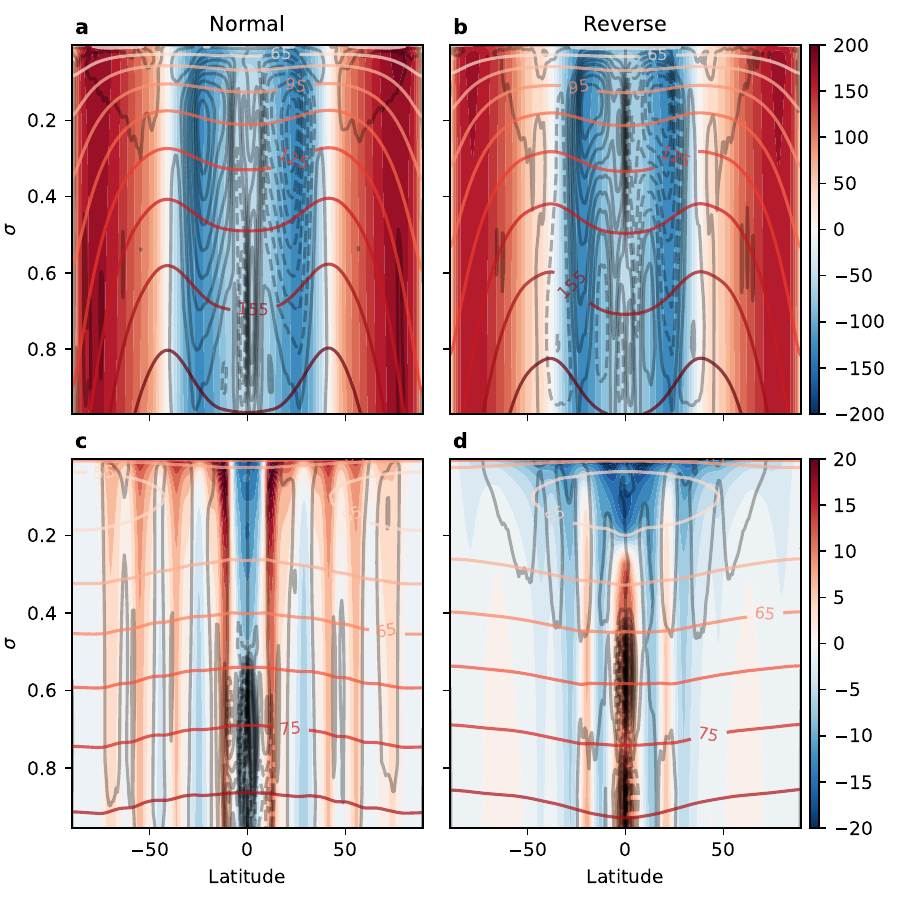}
    \caption{Comparison between the simulated flow using T$_{\rm{Normal}}$ and T$_{\rm{Reverse}}$ for 10 and 1 bar (top and bottom row, respectively). The shading is for the zonal mean zonal wind, black contours are for the zonal mean meridional streamfunction, and red contours are for the temperature. All simulations are conducted with T170 horizontal resolution.}
    \label{fig:temp_rn}
\end{figure*}

In this section, we describe how we tested the dependence of the simulated flow on the meridional structure of the forcing temperature $T_{\rm{eq}}$. We did this by changing the $\Delta T(\phi,\sigma)$ term in $T_{\rm{fit}}$ (Equation~\ref{eq:fit}) to change its meridional structure (Figure~\ref{fig:Teq}; see a detailed description in the Appendix, namely Equations~\ref{eq:DTgauss}-\ref{eq:DTcos}). First, we compared the simulation forced by $T_{\rm{Obs-like}}$ (Equation~\ref{eq:DTgauss}) and $T_{\rm{Obs}}$. The main difference between the two is a simplified vertical decay of the meridional temperature difference, neglecting a more complex vertical structure of the meridional temperature gradient. A comparison of the zonal mean zonal winds with the simulation forced with $T_{\rm{Obs}}$ and $T_{\rm{Obs-like}}$ (black and blue curves in Figure~\ref{fig:figusims}c, respectively), shows that for both the $10$ and $30$ bar simulations, the resulting winds have some qualitative differences between $T_{\rm{Obs}}$ and $T_{\rm{Obs-like}}$ (solid and dotted black and blue curves in Figure~\ref{fig:figusims}c). Using $T_{\rm{Obs-like}}$ results in weaker winds both at the equator and the midlatitudes. The midlatitude flow in this case is wider and has a unique latitudinal structure, with the winds peaking near the pole (solid and dashed blue curve in Figure~\ref{fig:figusims}c). The zonal winds in this idealized simulation are also qualitatively similar in the sense that the different simulations produce retrograde winds in the equatorial regions and prograde jets in the midlatitudes. Notably, the simulations forced with the idealized forcing produce jets that are more similar to the observed fit in that they do not produce a retrograde jet in the polar latitude (Figures\ref{fig:temp_rn}a-b, and ~\ref{fig:figusims}c).

The comparison between simulations forced with $T_{\rm{Obs}}$ and $T_{\rm{Obs-like}}$ points to the role of the vertical structure of the meridional temperature gradient; this difference is consistent for both 10 and 30 bar (for high enough resolution, Figure~\ref{fig:figusims}). A comparison between $T_{\rm{Obs-like}}$, $T_{\rm{Normal}}$, and $T_{\rm{Reverse}}$ aims to answer the question of what effect the meridional gradient of $T_{\rm{eq}}$ has on the resulting zonal winds. More specifically, this is motivated by the differences in the solar forcing on Uranus and Neptune. In the annual mean, Neptune receives more solar radiation at the equator and less at the poles (similarly to $T_{
\rm{Normal}}$) and Uranus receives more solar radiation at the poles and a minimum at the equator (similarly to $T_{\rm{Reverse}}$. As in the previous comparison, the answer depends on the depth of the forcing temperature's bottom layer and resolution (magenta curves in Figure~\ref{fig:res_dep}, and Figure~\ref{fig:figusims}c). At high enough resolution and deep enough simulation, the resulting winds are not sensitive to the imposed meridional gradient. This striking depth dependence is clear when comparing the top and bottom rows in Figure~\ref{fig:temp_rn}. The top row in Figure~\ref{fig:temp_rn} is for simulations with a 10 bar bottom layer, and shows that except for small quantitative differences between the normal and reverse cases, the simulated flow, streamfunction, and temperature for both the normal and reverse cases are close to identical. This is clearly not the case when the bottom layer is restricted to 1 bar (Figure~\ref{fig:temp_rn}). This result offers an explanation for the qualitatively similar winds observed in both Uranus and Neptune (Figure~\ref{fig:fig_obs}), despite their distinct thermal forcings, which point to the important role of the dynamics in the two planets.

\subsection{Horizontal resolution dependence}

\begin{figure}[htb!]
    \centering
    \includegraphics[width=\linewidth]{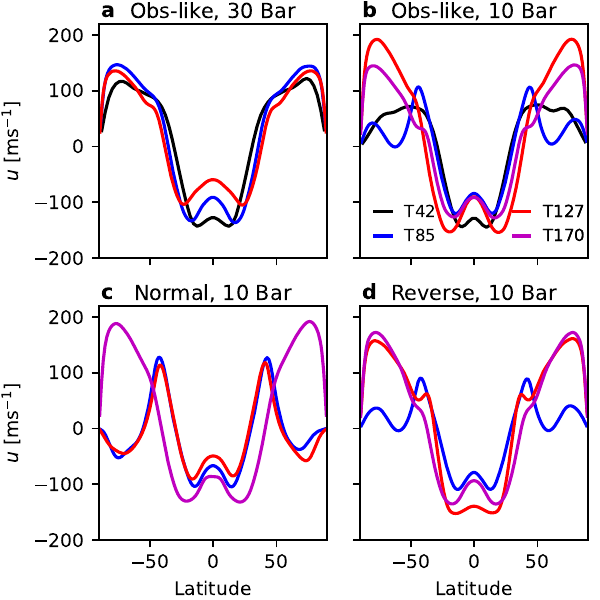}
    \caption{
    Zonal mean zonal wind (m s$^{-1}$) as a function of latitude for idealized $T_{\rm{eq}}$ experiments at different resolutions: T42 (black, panels a-b), T85 (blue), T127 (red), and T170 (magenta panels b-d). (a) 30 bar simulation with Obs-like $T_{\rm{eq}}$. (b-d) 10 bar simulations with Obs-like (b), Normal (c), and Reverse (d)  $T_{\rm{eq}}$.}
    \label{fig:res_dep}
\end{figure}

The results described above are resolution dependent for the 10 bar simulation; at lower resolutions there are differences between $T_{\rm{Normal}}$, $T_{\rm{Obs-like}}$, and $T_{\rm{Reverse}}$ (blue curve in Figure~\ref{fig:res_dep} panels b-d). We systematically tested this horizontal resolution dependence using the different idealized $T_{\rm{eq}}$. We find that the resolution sensitivity depends on the depth of the simulation and the meridional structure of the forcing temperature. For the 30 bar simulation, there is only a small resolution dependence, with the simulated flow being similar going from a coarse T42 ($2.8^{\circ}\times2.8^{\circ}$; black line in Figure~\ref{fig:res_dep}a) resolution to a finer T127 ($1^{\circ}\times1^{\circ}$; red line in Figure~\ref{fig:res_dep}a). In contrast, this is not the case for the 10 bar simulations, where there is a strong resolution dependence that also depends on the meridional structure of $T_{\rm{eq}}$. For the case of $T_{\rm{Obs-like}}$ the simulated flow differs between T42, T85 ($1.4^{\circ}\times1.4^{\circ}$), and T127 (black, blue, and red lines in Figure~\ref{fig:res_dep}b, respectively). This is not the case for T127 and T170 ($0.7^{\circ}\times0.7^{\circ}$) resolution, where the simulated flows are qualitatively similar (magenta and red lines in Figure~\ref{fig:res_dep}b). A similar resolution dependence is observed for the $T_{\rm{Reverse}}$ simulation (Figure~\ref{fig:res_dep}c). However, in the $T_{\rm{Normal}}$ simulation there is a qualitative difference in the simulated flow between using T127 and T170 (magenta and red lines in Figure~\ref{fig:res_dep}d).

This depth-dependent resolution sensitivity is suggestive of the role of the dynamics and more specifically the eddies in the resulting simulated flow. One possible hypothesis comes from considering the Rossby deformation radius $L_R = NH/f$ \citep{vallis_atmospheric_2017}. Given that $L_R\propto H$, a deeper simulation will result in a larger Rossby deformation radius, and thus larger eddies are more dominant. This explanation fits well with the small resolution dependence of the 30 bar simulations, the resolution dependence of the 10 bar simulation, and its detailed dependence on the temperature gradient, more specifically, the difference between the idealized $T_{\rm{eq}}$ at T127 and the similarity at T170 (Figure~\ref{fig:res_dep}). The differences in the meridional temperature gradients dictate the latitude of eddy generation, and therefore the relevant value of $f$ that contributes to $L_R$. Notably, at a high enough resolution the 10 bar simulations are also insensitive to the detailed merdional temperature gradient imposed in $T_{\rm{fit}}$ (Equation~\ref{eq:fit}), as are the 30 bar simulations (Figure~\ref{fig:figusims}), and both reproduce zonal winds that are qualitatively similar to the observed ones. This again underscores the importance of eddies in maintaining the circulation on these planets.

\subsection{Model parameter dependence}

\begin{figure*}[htb!]
    \centering
    \includegraphics[width=7in]{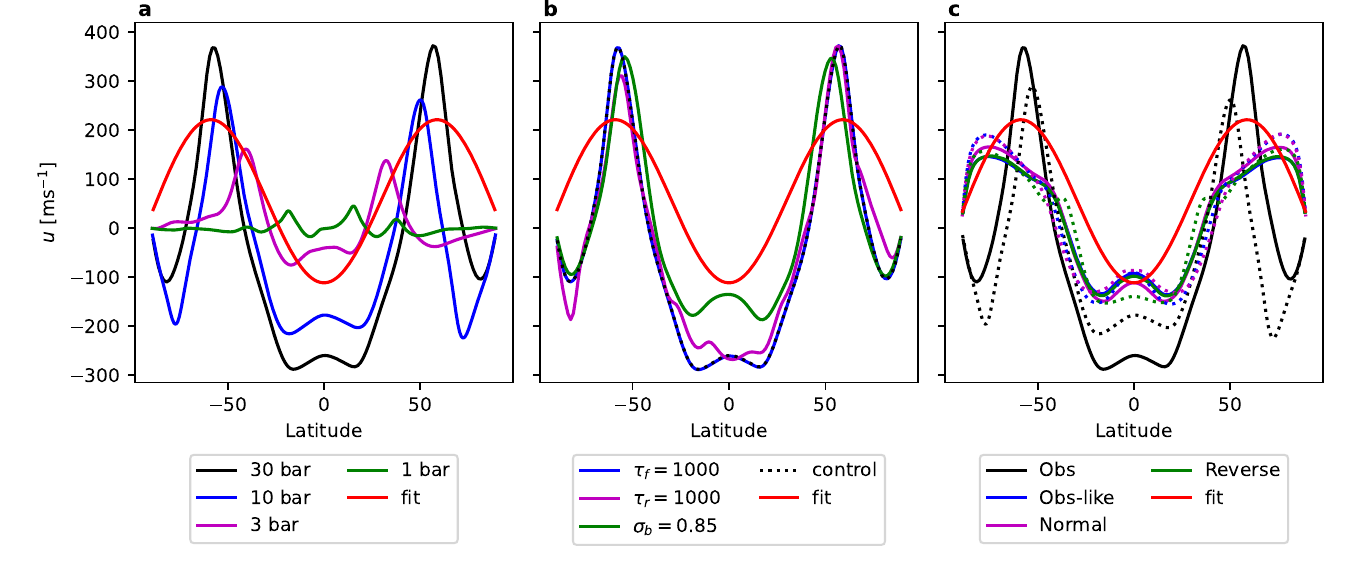}
\caption{Zonal mean zonal winds at $\sim250$ hPa for different model simulations. (a) Simulations with different bottom depths using $T_{\rm{Obs}}$ with resolutions detailed in Table~\ref{tab:res} (b) Simulations testing the sensitivity to different model parameters using $T_{\rm{Obs}}$ and 30 bar bottom layer depth. (c) Simulations with the idealized $T_{\rm{fit}}$ forcing, where the dotted lines are for simulations with a depth of 10 bar. In all the panels the red curve is the analytical fit for the observation by \citep{sromovsky_dynamics_2005}}
    \label{fig:figusims}
\end{figure*}

In this section, we describe how we tested the dependence of the simulated flow on different model parameters. Figure~\ref{fig:figusims}b shows the influence of some model parameters on the simulated zonal mean zonal wind. More specifically, changes in the thermal relaxation timescale ($\tau_r$) and the frictional timescale ($\tau_f$) have a small influence on the winds (magenta and blue curves, respectively), with the thermal relaxation timescale being more important. In addition to its stronger influence, a longer relaxation timescale results in a long time to reach a steady state; for example, for a simulation with $\tau_r=1000$ it takes $\approx35000$ simulated days to reach a steady state, compared with a few thousand days for all other simulations with $\tau_r=100$. A more significant response occurs when the depth of the boundary layer is increased ($\sigma_b$ from 0.97 to 0.85). In this case, the equatorial winds are weaker, and the midlatitude jets have a slight equatorward shift and some weakening. We note that it is possible that the effect of a longer frictional timescale ($\tau_f$) in a setting with a deeper boundary layer will be stronger than the effect shown here, but will not change the simulated winds qualitatively.

\section{Discussion}\label{sec:diss}
Uranus and Neptune are the least explored planets in the Solar System. As a result, observations are scarce and the understanding of different processes that occur in these planets is limited. The advocacy for orbiter missions to these planets has increased in the last decade. Simulating the circulation on these planets is thus important for the prediction and interpretation of current and future observations. Only limited studies have invested efforts to simulate the circulation on these planets. The result of these efforts emphasize the challenges of simulating the circulation on these planets.

More specifically, in both deep and shallow atmosphere models, it was challenging to simulate the observed wind on these planets, unless unrealistic forcing is imposed \citep{lian_generation_2010,aurnou_effects_2007}. In a recent effort \cite{milcareck_zonal_2024} have used a shallow atmosphere model with realistic radiation. Although their approach seems promising, their simulated flow seems different from the observed ones, both in magnitude and in meridional and vertical structure. For this study we used an idealized Newtonian cooling setting to test how different aspects of the forcing temperature affect the simulated circulation.

A main result of this study suggests that the simulated circulation has a strong dependence on the depth of the forcing temperature that we use. We see that as we extend the forcing temperature deeper, the zonal mean zonal winds become stronger and shift more poleward. A significant response occurs when we go to $\sim 10$ bar or deeper, where in addition to changes in the zonal mean zonal winds, the zonal mean meridional circulation becomes qualitatively different and has similar characteristics to that suggested by the observations, and for simulations with 30 bar atmosphere there is also a weak resolution dependence (Figure~\ref{fig:sim}). This complex circulation is maintained by meridional and vertical eddy fluxes (Figure~\ref{fig:figmom}).

A possible hypothesis that arises from our results testing the model dependence on the depth, model parameters, $T_{\rm{eq}}$, and resolution is that for a deep enough atmosphere, the model is no longer sensitive to details of the thermal forcing and the resulting flow is dictated by internal dynamics of the model. This result offers an explanation for the similarity of the flow observed on Uranus and Neptune despite the differences in their radiative and internal forcing. Additionally, this is an after-the-fact justification for the use of Uranus parameters and its observed temperature. That said, future studies are needed to verify this hypothesis. Our results also offer a partial explanation for the discrepancy between simulations with shallow models and observations, as most shallow models in existing literature are extended to only 3 bar \citep[e.g.,][]{liu_mechanisms_2010,milcareck_zonal_2024}. 

In addition to the lack of observations, simulating the circulation of the ice giants can become a computationally expensive task, especially as the realism of the GCM increases. One example is that as a result of the cold temperatures on the ice giants, a GCM with realistic radiation will require a long time to reach radiative equilibrium. This timescale will increase as the model simulates deeper parts of the atmosphere, as the thermal relaxation timescale increases with atmospheric mass. Including the representation of other processes, such as convective or different diabatic processes that are assumed to be important for ice giants \citep[e.g.,][]{hueso_convective_2020,guillot_condensation_1995,gierasch_vertical_1987}, increases the complexity and uncertainty of the GCM. However, the depth-dependent resolution sensitivity found in this study might suggest a potential compensation between conducting deeper simulations with a low resolution and a shallower one that will require high resolution. 

While taking into account different processes is a desirable approach to model the atmospheric circulation, this study shows the benefit of simplifying the system to understand the role of key processes and parameters. Additionally, idealized models are generally cheaper computationally, making them a good tool to use in the process of optimizing a realistic GCM for the ice giants, especially as simulating a deep atmosphere can be a numerically expensive task. In this study different processes are neglected; however, this modeling framework is flexible, and future work can use this framework to test how dry convection plays a role in the dynamics, can add simple representations of diabatic processes, or can force the same model with temperature profiles that are an output of a radiative convective equilibrium model.

\begin{acknowledgements}
We acknowledge support from the Israeli Space Agency and the Helen Kimmel Center for Planetary Science at the Weizmann Institute of Science. IG is supported by the Cooperative Institute for Modeling the Earth System. We thank Keren Duer for her help with producing Figure 1.
\end{acknowledgements}

\bibliographystyle{aa} 
\bibliography{Ice_Giants_clean} 

\begin{appendix}
\section{Fitted forcing temperature}\label{app:a}
The idealized forcing temperature takes the general form of $T_{\rm{fit}}=T_v(\sigma)+\Delta T(\phi,\sigma)$ (Equation~\ref{eq:fit}), where $T_v$ is for the vertical temperature profile and $\Delta T$ represents the meridional temperature structure. The vertical profile is given by
\begin{equation}
    T_v(\sigma) = \left(\left(T_0\sigma^{\alpha\kappa}\right)^{\beta}+\left(T_1\sigma^{\zeta}\right)^{\beta}\right)^{1/\beta},
\end{equation}
where $\kappa=C_p/R$ and $C_p$ is the heat capacity and $R$ is the gas constant, $\alpha$ is a coefficient that accounts for deviations for an adiabat, where in the case of Uranus and Neptune we use $\alpha=1.25$ to mimik the superadiabtic temperature profile \citep{guillot_condensation_1995}, $T_0,\ T_1,\ \zeta \ \rm{and} \ \beta$, are constants. This structure of the vertical temperature profile has similarities to the one used in \cite{showman_atmospheric_2019}, where the first term ($T_0$) represents the region in the atmosphere controlled by convection, and thus the temperatures increase with depth and the second term ($T_1$) controlled by radiation and temperatures increase with height (where the slope controlled by $\zeta$, and $\beta$ mostly acts to smooth out the profile).

The meridional temperature structure is different for different cases, for the one that aims to mimic the observations we use
\begin{equation}\label{eq:DTgauss}
    \Delta T = \delta T\left\{\exp\left[-\left(\frac{\phi-\phi_0}{\sigma_\phi}\right)^2\right]+\exp\left[-\left(\frac{\phi+\phi_0}{\sigma_\phi}\right)^2\right]\right\}D(\sigma),
\end{equation}
which is a linear combinations of two Gaussians centered at $\phi_0$, $35$ in our case, with a width of $\sigma_\phi$ with an amplitude of $\delta T$ multiplied by a decay function $D(\sigma)=c+a\exp\left(b\sqrt{\sigma}\right)$, where c, a, and b are constants. In this form, the decay is uniform in latitude. For the normal and reverse cases, 
\begin{equation}\label{eq:DTcos}
    \Delta T = -\delta T\cos\left(\phi-\phi_0\right)^2 D(\sigma)
\end{equation}
where $\phi_0=0$ and $\phi_0=90$ for the normal and reverse cases, respectively. These are the constant values that produce the different $T_{\rm{fit}}$ in Figure~\ref{fig:Teq}: $\sigma_\phi=20$, $c=0.05$, $a=1.5$, $b=-5$, $T_0=180$, $T_1=39$, $\zeta=0.06$, $\beta=5$. These parameters are used to fit to the observed profile with $p_s$ (in $\sigma=p/p_s$) being 10 bar and following that extended to 30 bar.

\section{Simulations convergence}\label{app:b}
Here we show the different timeseries of the zonal mean zonal winds at $\sim250$ hPa, for different resolutions and different idealized equilibrium temperatures for the 10 bar simulation.
\begin{figure*}[htb!]
\centering
    \includegraphics[width=\textwidth]{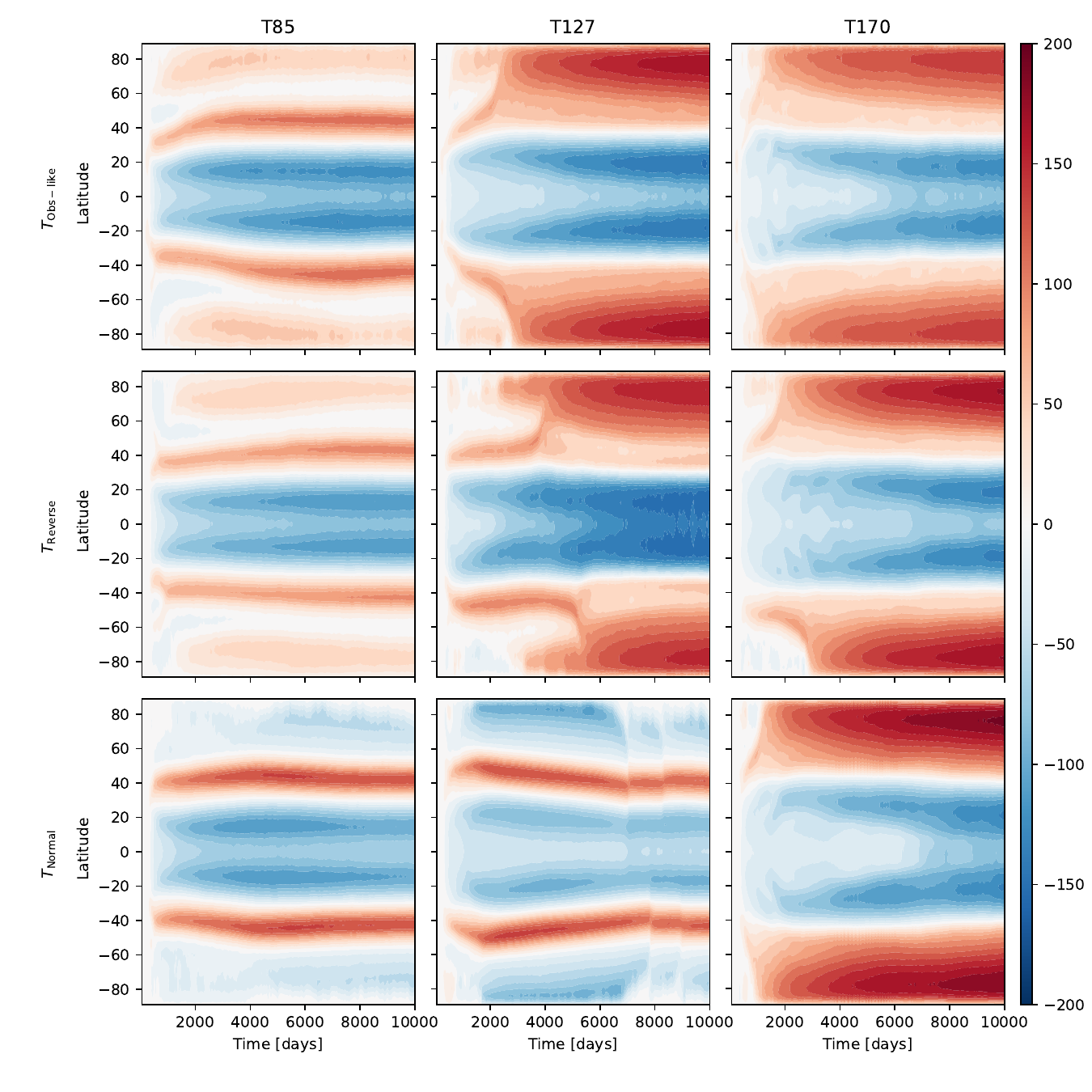}
    \caption{Time series of the zonal wind zonal winds (ms$^{-1}$) at $\sim250$ hPa for the first 10000 days of simulation for different resolutions (columns) and $T_{\rm{eq}}$ (rows).}
    \label{fig:conv}
\end{figure*}

\section{Momentum balance for 30 bar simulation}
\begin{figure*}[htb!]
    \centering
    \includegraphics[width=0.95\textwidth]{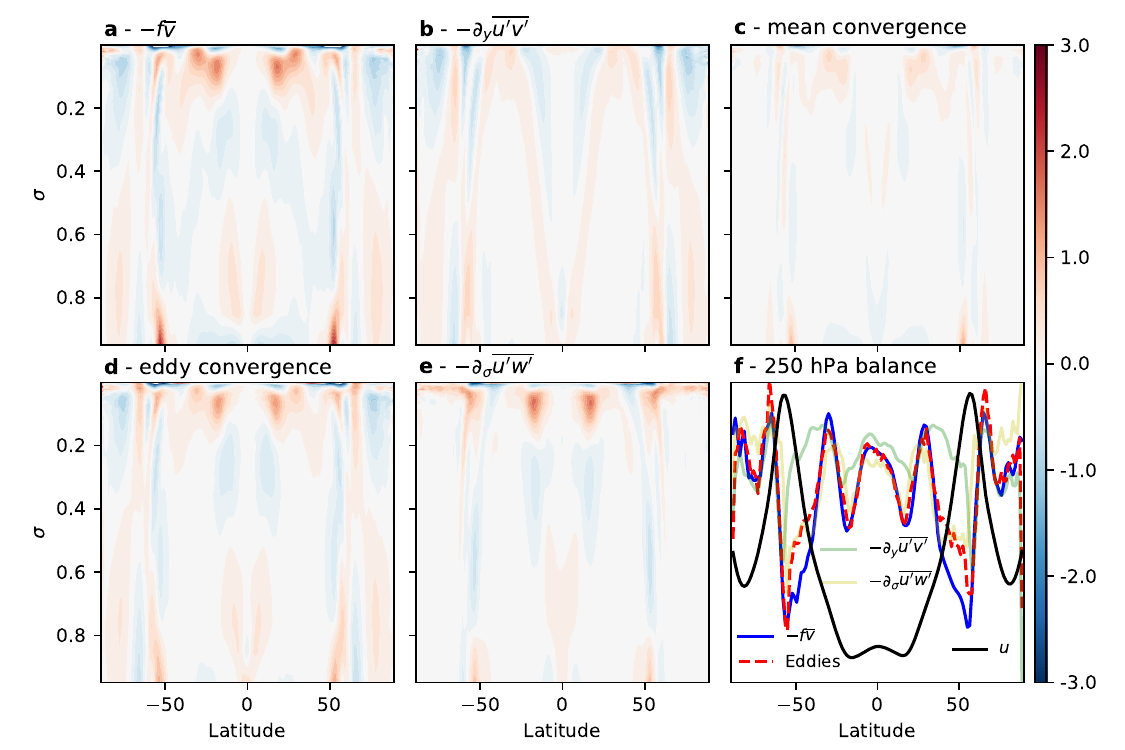}
    \caption{Similar to Figure~\ref{fig:figmom}, with the streamfunction contours in panel a, for a simulation with a 30 bar bottom layer}
    \label{fig:app_30}
\end{figure*}

\end{appendix}

\end{document}